 \documentclass[preprint,3p,times]{elsarticle}

\usepackage{amssymb}
\usepackage{setspace}
\journal{Nuclear Instruments and Methods in Physics Research}

\begin{document} 

\begin{frontmatter}

\title{A conceptual issue on the statistical determination of the neutrino velocity}

\author{Antonio Palazzo}
\address{ Cluster of Excellence, Origin and Structure of the Universe, Technische Universit\"{a}t M\"{u}nchen,\\
 Boltzmannstra\ss e 2, D-85748  Garching, Germany}

\begin{abstract}
We discuss a conceptual issue concerning the neutrino velocity measurement, 
in connection with the statistical method employed by the OPERA collaboration 
for the inference of the neutrino time of flight. 
We expound the theoretical framework that underlies the delicate statistical procedure 
illustrating its salient aspects. In particular, we show that the order of the two operations of sum 
and normalization used to combine the single waveforms so as to build the global PDF is a crucial point. 
We also illustrate how a consistency check able to test correctness of the PDF-composing
procedure should be designed.
\end{abstract}

\end{frontmatter}


\section{Introduction}

The OPERA collaboration has recently reported~\cite{OPERA} on a smaller time of flight 
of CNGS muon neutrinos  with respect to that expected assuming propagation
at the speed of light in vacuum.%
\footnote{Unless explicitly stated we always refer to the second version of the OPERA preprint.
The first version is mentioned only in the footnote (4) and in the final note added at the end of the paper.}
A few weeks ago the OPERA collaboration has identified two possible instrumental
effects that could have influenced its neutrino timing measurement. Furthermore, a few days ago,
an independent measurement performed by the ICARUS collaboration~\cite{ICARUS} has found no 
evidence of neutrino superluminal propagation, thus rejecting  the anomalous OPERA result. 
 
Notwithstanding, measuring with better precision the velocity of neutrinos remains an important  goal 
and other collaborations are already at work with this purpose.  In such a landscape, any issue of interest to the 
OPERA collaboration is inherently of interest to a larger part of the scientific community.
With this spirit, in this paper,  we address a conceptual issue pertaining the statistical 
procedure employed by the OPERA collaboration to infer the neutrino velocity.

While referring to the OPERA case for definiteness, our considerations will be valid for any kind of long-baseline setup, which
makes use of proton waveforms as neutrino pulsed sources. We stress that  our discussion
is independent of the width of the proton waveforms, and is relevant also for 
short pulses measurements as those performed in a second phase by the OPERA
collaboration and by ICARUS,  and which (presumably) will be adopted also by future experiments. As we will show
in detail, any kind of neutrino velocity measurement performed at a long-baseline detector
is intrinsically {\em a statistical measurement} which relies upon the conceptual framework
we are going to expound. 

\section{A basic question}

We remind the reader that the first  measurement performed by OPERA 
is based on a statistical comparison of the time distribution of the protons
ejected at CNGS (equivalent to that of the emitted neutrinos)  with that of the neutrinos observed
in the OPERA detector, where a total number $N \sim 15000$ of interactions have been recorded.  
The time distributions (waveforms) of protons have been measured for each $10.5\, \mu$s-long 
extraction for which neutrino interactions  are observed in the detector. Then, from their combination, 
the global probability density function (PDF) of the neutrino emission times is obtained.  Finally, 
such an emission PDF is compared with the time distribution of the detected neutrinos through  
standard maximum likelihood  analysis.

It should be noted that such a kind of procedure tacitly  assumes that it is possible to make 
a one-to-one correspondence between neutrino interactions and proton waveforms.
 Although perfectly
legitimate and very reasonable, such an assumption leads to conceptual 
consequences that must be taken into proper account
in the procedure itself. The main point is the following. 
The aforementioned one-to-one correspondence 
is formally equivalent to assume a certain degree of prior knowledge 
on the neutrino velocity.
With it, we are declaring of being certain%
\footnote {At a formal level, the declared prior degree of knowledge on the neutrino velocity $v$ corresponds
to impose that the conditional probability $P(B|A)$ of the event B given the event A is equal to one, where,
in the situation under study: The (conditioning) event A is represented by a neutrino interaction at a time $t_A$, and the
(conditioned) event B is the emission of such a neutrino by a waveform departed around the earlier time
$t_B = t_A - L/v$, from a source located at a known distance L from the neutrino interaction point. It is
worthwhile to underline the unusual (inverted) time-ordering of the two events A and B ($t_B<t_A$), 
which is responsible for what may seem counterintuitive conclusions.}
that  a given neutrino interaction has been produced by a neutrino emitted
by a  given proton extraction whose duration is about $10.5\, \mu$s.%
\footnote {Of course such a kind of one-to-one correspondence is unavoidable  also when using 
shorter neutrino bunches. In any case, given the time of a neutrino interaction in the detector,
one can identify the originating bunch (and discard the remaining ones) only making some assumption on the neutrino velocity.}
Now, the question naturally arises: Which is the weight  one has to attach to a 
given proton waveform in the global PDF?
In our opinion  it must be equal to the number $n_i$ of detected neutrinos associated 
to the corresponding ($i^{th}$) extraction and {\em be  independent} on 
the intensity of the waveform. Varying the index $i$, the number
$n_i$ is almost always zero,  sometimes it is one, and more rarely it is a bigger integer number. 
It is this integer number to inform us on how much a particular waveform effectively 
contributed  to determine the estimate of the neutrino time of flight, and not the 
intensity of that waveform (or any other weight factor proportional to it).

As an example, suppose that, because of a rare statistical fluctuation, we had
found a high number of detected neutrinos  (say  $n_{1} = 10$)  associated to a given
waveform $W_{1}$ having intensity $I_{1}$ not much different
from that of all the other waveforms.  This means that the shape information encoded by $W_{1}$  
is represented ten times in the time distribution of the $N$ detected neutrinos, independently
of its (particularly low) intensity $I_1$. As a further example, suppose that, again for a rare statistical
fluctuation, we had found that a given waveform $W_{2}$ having a smaller-than-average 
intensity $I_2$ (say one tenth of the ``normal'' intensity), all the same, 
has produced one neutrino interaction  ($n_{2} = 1$) in OPERA. Also in this case, what matters
is the number (one) of detected  neutrinos, and not the waveform intensity. 
Although having a very low intensity, the waveform $W_{2}$ 
has indeed provided an amount of information on the neutrino time of flight, which 
is equal to that furnished by any of those other waveforms that have produced, like $W_{2}$,  only one interaction.

A global PDF faithfully representative of {\em the emitted neutrinos associated to the detected ones} 
will contain only a (very) partial information on the fluctuations of the intensity of the individual waveforms.
Most of this information gets lost  in the poissonian neutrino detection process which, discretizing
the original information upon the waveform intensity, loses almost any memory of it. 
The only (partial) account of the intensity of a given waveform  $W_i$ is that provided by the number $n_i$
of detected  neutrinos associated to it. Accordingly,  when building the global PDF, 
one should {\em first} normalize to $n_i$ each single proton waveform and {\em then}  sum them together
(obtaining automatically, by construction, a sum equal to the total number $N$ of the observed events). 
The correct order of the two operations of sum and normalization is thus: {\em First  normalize and then sum}.%
\footnote{This is the opposite order with respect to that apparently advocated by the OPERA collaboration
in the first version of~\cite{OPERA}. See the note added at the end of our paper.} 

\section{A gedanken-experiment}

In order to elucidate the importance of the ordering of the two operations,
the following {\em gedanken-experiment} can be envisaged. Suppose that we start from the
hypothesis that the neutrino velocity is well known and has a true value $v_ {true}$.
Suppose also  that all the proton $10.5\, \mu$s-long waveforms are 
identical and represented by the positive definite function $g(t)$ having
unitary area. 
Note that the first hypothesis is of the same character of that made by OPERA: It
is only quantitatively stronger. Indeed, it just implies a more precise {\em a priori} 
association among neutrino interaction times and proton extraction times. The second
one renders the global PDF identical to the function $g(t)$ modulo a multiplicative
factor (the total number $N$ of detected neutrinos). Now, suppose that we are able to measure the
neutrino time of arrival with a precision $\delta t$ much smaller than
the waveform width.
Then, for each interacted neutrino we can identify the time when it has been emitted  
at a given distance L with precision $\delta t$. 
This means that within the three-years-long time series emitted at the CNGS, 
we can identify a small time interval of duration $\delta t$ during which the detected neutrino has been emitted. 
By construction, this will be always a sub-interval of the associated longer proton extraction.

This presumed knowledge entitles us  to retain only that sub-interval from the total longer extraction, 
discarding all the rest of it. By repeating such a procedure for all the neutrino interactions, we will obtain a 
series of $N$ $\delta t$-long time sub-waveforms. Each of them will be nothing else than a thin slice of the original
waveform centered around a time $t$ lying in the interval $[0,10.5]\mu$s, and having height equal to $g(t)$.
The $N$ sub-waveforms can be combined together as to obtain a new global probability density function. 
It is not difficult to realize that, if one combines the single sub-waveforms by first summing them together 
and then performing the normalization of their sum, a global PDF will be obtained,
which will be different from the original waveform function $g(t)$.  Indeed, with this method, 
at each  value of the time $t$ inside the  $10.5\, \mu$s-long  interval, the original waveform will
be counted (erroneously) two times:  A first counting factor comes from the fact that the 
number of times the small sampling sub-interval will lie  around $t$ is proportional to $g(t)$ itself;  
A second counting factor arises from the fact that, with such a method, the area under any sub-waveform 
[simply given by $\delta t \times g(t)$] will be proportional to $g(t)$, the height of the sub-waveform itself. 

In the limit $\delta t \to 0$ a PDF proportional to $g^2(t)$ will be generated. 
This is clearly  a paradoxical result: By hypothesis all the waveforms are identical 
and equal to the function $g(t)$ and therefore we must recover a PDF proportional to
$g(t)$ if our PDF-composing method is not biased.  It is not  difficult to recognize that
the only way to resolve the paradox and obtain a PDF proportional to $g(t)$ is to invert the order of the two 
operations of sum and normalization, performing {\em first the normalization and then the sum}. 
In this case, for any value of $t$, the original waveform is counted (correctly) only one time,
since the second of the counting factors mentioned above is now removed, as 
the area under each sub-waveform is normalized to one (and it is no more proportional 
to $g(t)$ as it occurs with the wrong method) before the summing procedure is performed.

\section{Effect of the mis-weighing procedure in the real setup}

Let us now try to gauge the effect the wrong procedure would produce in the real setup.
To this purpose, it is useful to make the following preliminary observations: (I) In the OPERA
setup the number of detected neutrinos $n_i$ associated to any given waveform $W_i$ can 
be assumed to be always equal to zero or one, as the probability a waveform originates
more than one event is extremely low. This implies that the number of relevant waveforms is identical 
to that ($N$) of the neutrino events;%
\footnote{While this circumstance simplifies the mathematical treatment presented below, its
generalization as to include the cases in which $n_i$ can be bigger than one is not difficult 
to attain. A result valid in such a more general case is provided in the footnote inserted 
before Eq.~(\ref{dt_cont}).}
(II) Each of the $N$ waveforms is embedded in a
time reference frame provided by its own digitation window [see also the discussion 
presented after Eq~(\ref{main_formula})]; 
(III)  Before the waveforms' summation is performed such $N$ reference frames 
are superimposed with their origins aligned. A this point one deals with one single time reference 
frame, common to all waveforms. In general, in such a reference frame, different waveforms will 
have different average times 
\begin{equation}
T_i = \frac{\int dt\, t\, W_i(t)}{\int dt\, W_i(t)} \,.
 \end{equation}
Furthermore, in general, one expects  different waveforms will have different intensities 
\begin{equation}
I_i = {\int dt\, W_i(t)} \,.
 \end{equation}
When using the correct composing method the intensities $I_i$'s  are irrelevant as all the waveforms are
assigned identical (unitary) weight in the PDF, which can be expressed as 
\begin{equation}
F(t) = \sum_{i=1}^N{\frac{W_i(t)}{I_i}}= \sum_{i=1}^N{w_i(t)}\,,
 \end{equation}
where we have introduced the auxiliary waveforms 
\begin{equation}
w_i(t) \equiv \frac{W_i(t)}{I_i} = \frac{W_i(t)}{\int dt\, W_i(t)}  \,,
 \end{equation}
having unitary normalization.  In this case, the global PDF is automatically normalized
to the total number $N$ of detected neutrinos. When using the alternative wrong method, 
each waveforms is assigned a weight factor $a_i$ proportional to is intensity $I_i$, 
and the PDF can be expressed as  
\begin{equation}
F'(t) = \sum_i^N{W_i(t)} = \sum_i^N{a_i w_i(t)}\,.
 \end{equation}
\begin{figure}[t!]
\vspace*{-2.45cm}
\hspace*{3.6cm}
\includegraphics[width=12.0 cm]{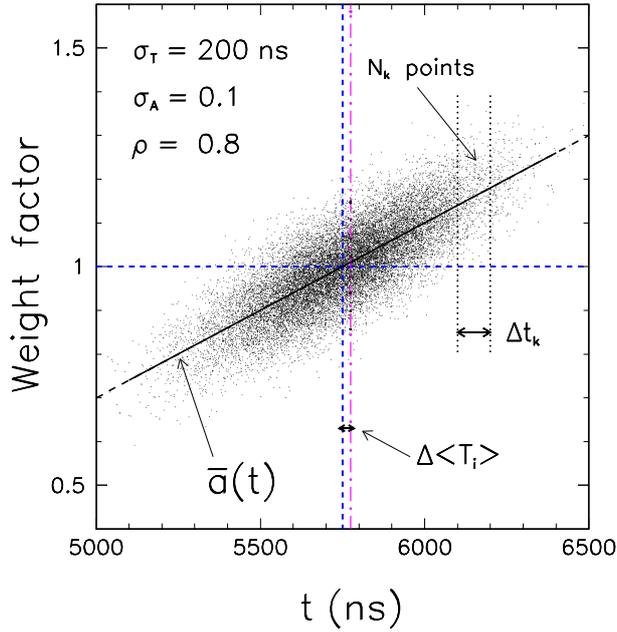}
\vspace*{-1.0cm}
\caption{Toy scatter plot of the amplitudes versus the mean times of 15000 waveforms. 
The points  have been extracted from  a bivariate normal distribution with
parameters indicated in the plot. The band delimited by the two vertical dotted 
lines indicates a generic time interval  $\Delta t_k$ centered around $\bar T_k$,
which contains $N_k$ points [see the discussion of Eq. (10)]. The dashed (dotted-dashed) vertical line 
indicates the average time obtained with the correct (incorrect) PDF-composing procedure.
The difference between of the two averages is $\Delta \langle T_i \rangle \sim 18\,$ns. See the
text for details. 
\label{fig_corr}}
\end{figure}  
In this case the correct normalization of the PDF is obtained by imposing the condition
\begin{equation}
\sum_{i=1}^{N} a_i = N\,,
\label{norm_amplitude}
 \end{equation}
equivalent to set equal to one the average value of the amplitudes ($\langle a_i \rangle = 1$). 
The  average time of the correct PDF is
\begin{equation}
\langle T_i\rangle = \frac{1}{N}\sum_{i=1}^{N} T_i \,,
 \end{equation}
while that obtained with the wrong PDF is
\begin{equation}
\langle T_i\rangle'  = \frac{1}{N}\sum_{i=1}^{N} a_i  T_i \,,
 \end{equation}
the difference among the two estimates being
\begin{equation}
\Delta \langle T_i \rangle \equiv \langle T_i \rangle' - \langle T_i\rangle =  \frac{1}{N}\sum_{i=1}^{N} (a_i -1)  T_i \,. 
\end{equation}
One can always time-order the $N$  times $ T_i $'s and divide the time
interval $[T_1,T_N]$ in an integer number $K$ of subintervals $\Delta t_k$'s of equal-width 
$\Delta t$. In this way the time shift takes the approximate form
\begin{equation}
\Delta \langle T_i \rangle \simeq  \frac{1}{N}   \sum_{k=1}^{K} (\bar a_k -1) \bar T_k N_k\,,
\label{dt_discrete}
\end{equation}
where $N_k$ is the number of waveforms having mean time
lying within the $k$-th subinterval positioned around their average time
$\bar T_k$, and obeying the normalization condition
\begin{equation}
\sum_{k=1}^{K} N_k = N\,,
\end{equation}
while $\bar a_k$ designates the average of their weight factors. The procedure outlined above can be
better appreciated with the help of Fig.~1, which displays  a toy scatter plot  of 15000 points, 
each one representing a waveform  with average time $T_i$ and amplitude $a_i$. 
In the limit $N\to \infty$ it makes sense to consider the limit $\Delta t \to 0$,
which allows us to establish the following correspondences   
\begin{eqnarray}
N_k \to& \frac{dN}{dt} dt\,,\\
\label{correspondence_n}
\bar a_k \to& \bar a(t)\,. 
\label{correspondence_a}
\end{eqnarray}
The density of points $dN/dt$ can be identified (apart from a proportionality factor) with 
the statistical distribution [call it $f(t)$] of the waveforms' mean-time variable $t$, while $\bar a(t)$ represents
the conditional expectation value  of the amplitude variable $a$ at a given value of the
variable $t$. In the continuos limit  the discrete sums in Eq.~(\ref{dt_discrete}) are 
replaced by integrals, thus obtaining  for the time shift%
\footnote{In the general case, in which more than one interaction per waveform
is present, it can be shown that Eq.~(\ref{dt_cont}) generalizes as follows. 
The integral at the numerator is replaced by the weighted sum of integrals  $\sum n p_n I_n$,
where $p_n$ is the probability that a waveform originates a number $n$ of neutrino events
and $I_n = \int{dt \,  [\bar a_n (t)-1]\, t\, f_n(t)}$.  The functions $f_n(t)$ are the statistical
distributions (normalized to unity) of the mean times of the waveforms associated to the $n$-tuple
of events.  The functions $\bar a_n(t)$ are the conditional expectation values of the amplitudes 
of the same waveforms at a given mean time t.  By construction, being $\sum p_n =1$, the denominator  $\sum p_n \int f_n(t) dt $ is 
unitary. In the OPERA setup, we estimate $p_2 \sim 10^{-5}$, so the ``higher-order'' terms beyond
the  first one ($I_1$) accounted for in Eq.~(\ref{dt_cont})  give a negligible contribution.}  
\begin{equation}
\Delta\langle T \rangle \equiv \lim_{\Delta t \to 0} \Delta \langle T_i \rangle =  \frac{\int{dt \,  [\bar a (t)-1]\, t\, f(t)}}{\int dt\, f(t)}\,.
\label{dt_cont}
\end{equation}
In order to proceed to the evaluation of the integrals in Eq.~(\ref{dt_cont}) we
must make some (reasonable) assumptions on the form of the statistical 
distribution of the mean times $T_i$'s and of the associated amplitudes $a_i$'s.
As a working hypothesis, it seems plausible to assume that their values are extracted 
 from a bivariate normal distribution,%
\footnote{To be precise, at a conceptual level, it would be more correct to make such kind of  
assumptions at the level of the distribution of $N$ waveforms {\em randomly
extracted} from the three-year-long train made of million of  waveforms
and then extract from this one the distribution of the $N$ waveforms associated to
the detected neutrinos. Indeed, these last ones constitute a biased sample of the original 
distribution, as the neutrino events tend to select  the most intense waveforms, being the probability
to originate a neutrino event  proportional to the waveform intensity.
If $G(a)$ is the native distribution of the waveforms' intensities, that  one sampled by the
neutrino events  will be $a\, G(a)$. In the case under study, the amplitude $a$ varies
only a few around unity and such a bias introduces only a small distortion from the
native distribution, entitling us to neglect such an effect.}
characterized by the two standard deviations $\sigma_T$
and $\sigma_A$ and by the correlation coefficient $\rho$. In such a case,
the following linear relation will hold  
\begin{equation}
\bar a(t)  = \langle a \rangle +  \beta (t - \langle t \rangle )\,,
\label{lin1}
\end{equation}
where  $\langle t \rangle$ and $\langle a \rangle$ designate
the averages of the two random variables and the coefficient 
$\beta$ is given by%
\footnote{We remind the reader that two bivariate normal random variables $X$ and $Y$
(also said to be ``jointly gaussian''), enjoy the property $E(Y|X) = E(Y) + \rho \frac{\sigma_Y}{\sigma_X} [X-E(X)]$, 
where $E(Y|X)$ is the conditional  expectation of $Y$ given $X$.} 
\begin{equation}
\beta = \rho \frac{\sigma_A}{\sigma_T}\,.
\label{beta}
\end{equation}
Taking into account that  the average amplitude  [see~Eq.~(\ref{norm_amplitude})]
is unitary by construction, Eq.~({\ref{lin1}) becomes 
\begin{equation}
\bar a(t)  = 1 +  \rho \frac{\sigma_A}{\sigma_T} (t - \langle t \rangle ) \,.
\label{lin2}
\end{equation}
Such a linear relation is represented by the regression line shown in Fig.~1.
The distribution $f(t)$ in Eq.~(\ref{dt_cont}) is by construction a gaussian distribution 
with standard deviation $\sigma_T$,
being a marginal distribution of the native bivariate normal distribution. 
By substituting Eq.~(\ref{lin2}) in Eq.~(\ref{dt_cont}), and  making use of elementary gaussian integrals,
one finally arrives at the result
\begin{equation}
\Delta \langle T \rangle = \beta\, \sigma_T^2 = \rho\, \sigma_A \,\sigma_T\,.
\label{main_formula}
\end{equation}
\begin{figure}[t!]
\vspace*{-2.7cm}
\hspace*{3.5cm}
\includegraphics[width=12.6 cm]{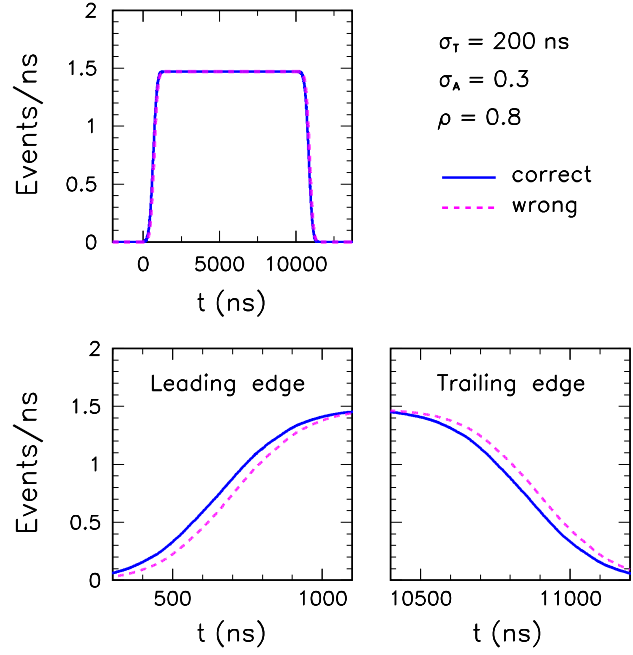}
\vspace*{-1.3cm}
\caption{Effect of the incorrect procedure of PDF-composing. Both PDFs
have been obtained summing up 15000 simulated rectangular waveforms having 
a fixed width of $10.2\,\mu$s. In one (correct) case the mean times
are random numbers extracted from a gaussian distribution with standard deviation $\sigma_T$,
while the weight factors are identical and unitary. In the other (wrong) case    
both the mean times and the weight factors are random numbers extracted
from a bivariate normal distribution with standard deviations $\sigma_T$ and
$\sigma_A$ and a correlation coefficient $\rho$. 
In the lower panels, representing a zoom of the edges of the two PDFs, a shift
$\Delta \langle T_i \rangle \sim 48\, $ns of the average time can be appreciated. 
The dispersion of the amplitudes $\sigma_A$ has been taken three times bigger
than its real estimated value for a better visual clarity.
\label{fig_pdf}}
\end{figure}  
Equation~(\ref{main_formula}) makes explicit what could be grasped on
an intuitive basis:
In the presence of a non-zero correlation among the waveforms' mean times
and their intensities, the use of the wrong procedure of PDF-composing leads
to a shift of the average time of the PDF.
In particular, for a positive (negative) correlation coefficient the wrong method
leads to an over(under)-estimation of the average neutrino emission time,
and a consequent under(over)-estimation of its time of flight.
Figure~2 illustrates such an effect on a toy PDF generated by
combining $15000$ rectangular waveforms of equal width,
and having mean times and intensities distributed according 
to a bivariate normal distribution. In this simple case, the shift of the 
average PDF time merely manifests itself as a shift of its edges.%
\footnote{In a more realistic situation, the mis-weighing procedure will give rise also to 
differences in the shape of the rest of the PDF. However, our toy model should
suffice to describe the salient aspects of the OPERA measurement, which is essentially
sensitive to the position of the two PDF's  edges.}

Let us now try to quantify the three parameters entering Eq.~(\ref{main_formula}):
(I) A reliable estimate of the dispersion of the intensities can be deduced from
the documentation publicly available on the CNGS website~\cite{CNGS}.
From Fig.~3 in~\cite{int_2009} and the two figures shown at page 24 in~\cite{int_2010}), 
we can infer $\sigma_A \sim 0.1$, at least for what concerns the 2009 and 2010 operational periods;%
\footnote{Such plots evidence a markedly asymmetric intensities' distribution, which presents
a second peak at intensities much lower than their average value. Therefore, appreciable corrections
to the gaussian approximation at the basis of our estimates are expected.  It is hard to gauge them 
without knowledge of the mean times of the associated waveforms.}   
(II) Without knowledge of the 15000 waveforms it is difficult to make a reliable estimate
of their mean times' dispersion  $\sigma_T$. An upper bound for this parameter can be
derived by observing that the width of  the PDF is bigger than that of the individual waveforms 
as a result of the broadening effect induced by the summing procedure. From Fig.~14 in~\cite{OPERA} a PDF width of about
$11.2\,\mu$s can be estimated, about $700\,$ns bigger than the width of the 
single waveforms. Assuming that the whole PDF enlargement derives from the dispersion of the 
mean times of the composing waveforms, we derive  $\sigma_T \sim 200\,$ns. 
The reliability of  such an estimate  is corroborated by the toy Monte Carlo simulations
used to produce the PDFs in Fig.~2, whose edges look quite similar to those
obtained in the real case (see Fig.~14 in~\cite{OPERA}).  Of course, part of the broadening effect
may come from other factors such as a variable width of the waveforms. Therefore,
such an estimate must be intended as an upper bound;
(III) Concerning the correlation coefficient $\rho$, we observe that for each waveform $W_i$, the start-time of the
digitation window is the trigger-time of the kicker magnet (see Fig.~3 and the related discussion on  pages 5 and 12 in~\cite{OPERA}).  
Such a time is set by the Wave Form Digitizer (WFD) as the common (arbitrary) time origin for each waveform before 
the summation procedure. This implies that shorter (longer) waveforms will have a smaller (bigger) mean time. 
In turn, the discussion made in~\cite{Gaxiola} (see the comment on Fig.~2 at page 4),
reveals that more (less) intense waveforms have systematically a longer (shorter) duration. 
Therefore, it seems quite natural to expect a positive correlation coefficient $\rho$,
although no firm conclusion can be traced  without having direct knowledge of the 15000 waveforms.

Inserting in Eq.~(\ref{main_formula}) the estimated value of $\sigma_A$ and 
the upper bound for that of $\sigma_T$  we deduce that,  for a positive correlation $\rho \sim 1$, 
a maximal shift of about  $20\,$ns can be induced. Such an upper bound  can be
slightly altered as a consequence of the following two factors:  
(I) A non-gaussian behavior of the intensities' and mean times' distributions;
(II) The potential effects (not considered in our toy simulations) of the wrong 
PDF-composing procedure on the shape of the flat-top of the PDF. 

\section{Relevance of the PDF-composing procedure to the future high precision short bunch measurements}

Although we have discussed the case of long waveforms
as those used in the first measurement by the OPERA collaboration, it is important
to stress that our considerations are valid independently of the duration of the 
proton pulses used to inject the neutrinos. In particular, our main result provided 
in Eq.~(\ref{main_formula}) remains valid for setups using very short bunches as those adopted
by the OPERA collaboration for their second cross-check measurement and by ICARUS, and 
which (presumably) will be used by future high precision experiments. 

One may be induced to think that 
the PDF approach is unnecessary when using very short pulses. However, this is not
the case if one intends to obtain a precision on the neutrino time of flight comparable 
or smaller than the width of the bunches themselves, which are typically a few nanoseconds long. 
To this regard it is important to observe that in the case of OPERA and ICARUS the short-bunch measurements 
were performed with the sole purpose of (dis-)confirming the (quite large) shift of $\sim60\,$ns  found in the first 
OPERA measurement, which is much bigger than the width of the short bunches (3$\,$ns). 
For such a specific purpose, it was sufficient to generate a histogram of the neutrino interaction
times (like that presented in Fig.~18 of~\cite{OPERA} or in Fig.~3 of~\cite{ICARUS} ) and check  that it was
statistically compatible or incompatible  with the original shift of about  $60\,$ns. 

However, the situation would be
completely different should one intend to use the short-bunch technique to measure time shifts (potentially) smaller than
the bunch width itself. In this case, resorting to the PDF method appears to be inescapable,
as it constitutes the only way to make a quantitative and precise comparison between the time distribution of  
the neutrino events and the detailed time structure of the emitted bunches.  In such circumstances implementing
the correct PDF-composing procedure will be essential.

\section{A consistency test}
 
In principle, a Monte Carlo test should allow the identification 
and the quantification of a problem in the procedure of PDF composing. However, we must note that 
this was not the case for the type of simulations performed by the OPERA collaboration. Indeed, according to the information 
reported in~\cite{OPERA} (see page 19),  and the more detailed documentation provided in the PhD thesis~\cite{tesi}
(see page 132), it emerges that the correctness of the PDF has been assumed {\em a priori}.
In fact, at page 19 of~\cite{OPERA} one can read: ``Starting from the experimental PDF, an ensemble of 100 data 
sets of OPERA neutrino interactions was simulated. Data were shifted in time by a constant quantity, hence faking
a time of flight deviation''. This implies that the correctness of the PDF was assumed {\em a priori} and was not tested
{\em a posteriori}. 

This circumstance lead us to deem it useful to show how a consistency check may be performed in order to test 
the correctness PDF-composing procedure. In addition, we note that although we think we have clearly shown 
which of the two methods is the correct one at the theoretical level, the reader may find useful
to have at his disposal also a numerical test of the issue in question. 
For such a test one has to reproduce a high number of  virtual 
experiments as follows. In each of them one should: (I) Propagate at the speed
of light the few-years-long train made of millions of waveforms so as to generate
$N\sim O(10^4)$ neutrino interactions in the detector with their own times;%
\footnote{It is conceptually insightful to observe that, in a faithful realization
of the real experiment, one should always make propagate the individual waveforms
and not the PDF as a whole, as done in the Monte Carlo test performed by the OPERA  collaboration.
In fact, the PDF should be thought as a mere statistical tool, resulting
from a {\em mathematical} sum of waveforms. In general, the PDF cannot be interpreted as 
a unique big physical neutrino wave as if it were originated from a {\em physical} 
sum of smaller impulses. If done, such a conceptually wrong interpretation would  ignore the
intrinsic poissonian nature of the detection process, leading to erroneous conclusions.}
(II)  Identify the associated $N$ waveforms;
(III) Build a toy PDF using one of the two alternative methods
we have discussed (first sum and then normalize and vice-versa) ; 
(IV) Compare such two PDF's with the distribution of the neutrino interactions times
so as to derive an {\em a posteriori} estimate of the neutrino velocity. By repeating
such a simulation a sufficiently high number of times, the proposed test
will tell the experimenters if the used PDF-composing method is correct
or not.  Indeed, when using the correct PDF-composing procedure, the estimate of 
the neutrino velocity obtained {\em a posteriori} will turn out to be statistically compatible with 
that imposed {\em a priori} (the velocity of light). When using the wrong procedure, 
two possibilities may occur: (A)  The {\em a posteriori}  estimate of the neutrino velocity 
is different from its value assumed {\em a priori}, thus indicating that the adopted 
method is wrong and making possible to quantify the error it has induced; (B) 
The {\em a posteriori}  estimate is compatible with that used {\em a priori}, thus indicating that the use of the wrong method
has harmless consequences: Although wrong at a conceptual level, it induces
an error that is quantitatively irrelevant. This second circumstance is what would occur 
in the presence of a negligibly small value of one (or more) of the three parameters
($\rho$, $\sigma_A$,  $\sigma_T$) entering Eq.~(\ref{main_formula}).%

\section{Conclusions}

We have discussed a conceptual issue concerning the  statistical measurement of the
neutrino velocity. We have evidenced that: 
(I) There are in principle two ways to obtain the 
global PDF from the single waveforms (first sum and then normalize or vice versa);
(II) The second is the correct one at a conceptual level;
(III) The use of the incorrect method  can lead to a wrong inference of the neutrino time of flight, an effect for
which we have provided  an analytical description; 
(IV) The Monte Carlo tests performed by OPERA would have been unable to 
identify a problem in the procedure of combination of the waveforms;
(V) We have proposed a consistency check able to
detect such a kind of problems.

\section*{Note added}

The original motivation of our paper was provided by the observation that the OPERA collaboration originally 
had reported an incorrect statistical procedure in the first version of~\cite{OPERA}. Indeed, reading that first version, it emerged that the 
single waveforms had been {\em first}  summed together and {\em then} their sum had been normalized
to the total number of neutrino interactions observed in the detector (see page 14 of the first version of~\cite{OPERA}).
The more extensive information provided in the PhD thesis~\cite{tesi}  (explicitly mentioned in~\cite{OPERA}) corroborated 
that circumstance (see pages 124-125 of~\cite{tesi}). 

After our preprint appeared, a second version of the OPERA paper~\cite{OPERA} has been posted on arXiv. 
Differently from the first one, the second version reports the correct PDF-composing procedure. 
The opposite and correct order of the two operations of summing and normalization is  now 
mentioned (see page 17 of the second version of~\cite{OPERA}), although no comment has been included to explain the
 change made on such a delicate point.  At present no further investigation on the issue in
 question is possible from outside the OPERA collaboration, as the relevant raw data (the time stamps
 of  the 15000 neutrino interactions and those of  the associated proton waveforms)
  have not been made publicly available.

\section*{Acknowledgments}
Our work  is supported by the DFG
Cluster of Excellence on  the ``Origin and Structure of the Universe''.


\end{document}